\documentclass{article}

\usepackage{PRIMEarxiv}

\usepackage[utf8]{inputenc} % allow utf-8 input
\usepackage[T1]{fontenc}    % use 8-bit T1 fonts
\usepackage{hyperref}       % hyperlinks
\usepackage{url}            % simple URL typesetting
\usepackage{booktabs}       % professional-quality tables
\usepackage{amsfonts}       % blackboard math symbols
\usepackage{nicefrac}       % compact symbols for 1/2, etc.
\usepackage{microtype}      % microtypography
\usepackage{lipsum}
\usepackage{fancyhdr}       % header
\usepackage{graphicx}       % graphics
\graphicspath{{media/}}     % organize your images and other figures under media/ folder
\usepackage{float}
\usepackage{dcolumn}% Align table columns on decimal point
\usepackage{bm}
\usepackage{amsmath}

\title{Predicting unavailable parameters from existing velocity fields of turbulent flows using a GAN-based model}
\author{
 Linqi Yu$^{1,*}$, Mustafa Z. Yousif$^{1, 2, *}$ \\
\And
 Young-Woo Lee$^{1}$, Xiaojue Zhu$^{3}$, Meng Zhang$^{1}$, Paraskovia Kolesova$^{1}$\\
\And
 Hee-Chang Lim$^{1, \dagger}$ \\
\\
 $^{1}$ School of Mechanical Engineering, Pusan National University\\
 2, Busandaehak-ro 63beon-gil, Geumjeong-gu, Busan 46241, Republic of Korea\\ \\
 $^{2}$ German Engineering Research and Development Center, LSTME-Busan Branch\\\\
 $^{3}$ Max Planck Institute for Solar System Research, Justus-von-Liebig-Weg 3, Göttingen 37077, Germany \\\\
 $^{*}$ These authors contributed equally to this work.\\
 $^{\dagger}$ Corresponding author, {\bf hclim@pusan.ac.kr}
}

\date{\today}% It is always \today, today,
\begin{document}
\maketitle

\begin{abstract}
In this study, an efficient deep-learning model is developed to predict unavailable parameters, e.g., streamwise velocity, temperature, and pressure from available velocity components. This model, termed mapping generative adversarial network (M-GAN), consists of a label information generator (LIG) and an enhanced super-resolution generative adversarial network (ESRGAN). LIG can generate label information helping the model to predict different parameters. The GAN-based model receives the label information from LIG and existing velocity data to generate the unavailable parameters. Two-dimensional (2D) Rayleigh-B{\'e}nard flow and turbulent channel flow are used to evaluate the performance of M-GAN. Firstly, M-GAN is trained and evaluated by 2D direct numerical simulation (DNS) data of a Rayleigh-B{\'e}nard flow. From the results, it can be shown that M-GAN can predict temperature distribution from the two-dimensional velocities. Furthermore, DNS data of turbulent channel flow at two different friction Reynolds numbers $Re_\tau$ = 180 and 550 are applied simultaneously to train the M-GAN and examine its predicting ability for the pressure fields and the streamwise velocity from the other two velocity components. The instantaneous and statistical results of the predicted data agree well with the DNS data, even for the flow at $Re_\tau$ = 395, indicating that M-GAN can be trained to learn the mapping function of the unknown fields with good interpolation capability. 
\end{abstract}

% keywords can be removed
%\keywords{Flow field denoising \and PIV \and Physics-guided DRL \and Machine learning}

\maketitle
\section{Introduction}\label{sec:introduction}
As a traditional topic in experimental and computational fluid dynamics (CFD), turbulence has been investigated extensively to enhance the understanding of its complicated characteristics and chaotic behavior over a hundred years. Visualizing and statistically analyzing turbulent flows requires massive data from flow fields. With various methods developed, high-fidelity turbulence data can be efficiently generated for aerodynamic simulations and other scientific applications. For instance, direct numerical simulation (DNS) has been extensively applied in the CFD field to simulate a variety of turbulent flows precisely at a specific range of Reynold numbers ($Re$), whose main mechanism solves the Navier-Stokes equations numerically \cite{Moin&Mahesh1998, Kimetal1987, Moseretal1999}. From the perspective of experimental measurement, one of the most widely-used methods is particle-image velocimetry (PIV), which plays a crucial role in the experimental investigation of turbulent flow fields \cite{Oudheusden2013}. However, if an ordinary two-dimensional (2D) PIV system is employed, only two velocity components, e.g., wall-normal velocity ($v$) and spanwise velocity ($w$), can be obtained but without streamwise velocity ($u$).

Note here that an unavailable parameter generally refers to a parameter that cannot be measured directly using 2D measurement techniques such as PIV measurement. For instance, if the normal section to the streamwise direction of the flow is measured, the streamwise component cannot be obtained directly and therefore is considered an unavailable parameter. In addition, pressure ($p$) and temperature ($T$) also cannot be obtained directly by those measurements. These unavailable parameters are undoubtedly important for turbulent flow analysis and engineering applications. Therefore, these three velocity components are required to analyze the instantaneous flow structures and statistics of turbulent flow \cite{Vanderweletal2019, Pope2000}. In addition, the fluctuating pressure field over the bluff bodies under turbulent boundary layers causes severe noise and vibration \cite{Blake2017}. Moreover, the pressure fields within the turbulent flow field are crucial to analyze the coupling mechanisms of turbulence–acoustic \cite{Manohaetal2000} and flow control strategies for reducing noise \cite{Eltaweeletal2014}. Temperature is also crucial in turbulence fields, as its gradient drives the heat transfer phenomenon. Additionally, temperature gradients in the near-wall region significantly impact the overall performance of many engineering systems \cite{Toutant&Bataille2013, Zhuetal2018}.

Currently, several methods are applied to obtain the aforementioned unavailable parameters. Therefore, in general, acquiring the full components of turbulent flow fields requires a high-performance PIV system such as a tomographic PIV (Tomo PIV) \cite{Scarano2012}, which realizes the acquisition of three-dimensional (3D) turbulent flow with full velocity components. However, in order to obtain the full velocity components, it would be necessary to incur additional costs and employ complex processes to achieve the desired level of precision. As for pressure measurement by PIV, Oudheusden \cite{Oudheusden2013} proposed a way how to manipulate the instantaneous pressure field by combining the experimental data with the governing equations, i.e., the Poisson equation. He {\it et al.} \cite{Heetal2020} used data assimilation to determine the pressure of turbulent velocity fields measured by PIV based on the unsteady adjoint formulation. However, implementing these methods still needs the assistance of Tomo PIV. In other words, the turbulent flow data from the 2D PIV system are insufficient to solve the governing equation or apply data assimilation because of the lack of one of the velocity components. Thermographic PIV (Thermo PIV) \cite{Allison&Gillies1997, Abrametal2018} has been developed to measure velocity fields and temperature based on thermographic phosphor particles, which possess temperature-dependent luminescence properties. Compared to the original PIV, Thermo PIV needs more extra setup and equipment, which include an extra ultraviolet laser and two extra cameras with suitable spectral filters. Nonetheless, it is difficult to predict unavailable parameters like $u$, $p$, and $T$, based on the available data ($v$ and $w$) obtained from an ordinary 2D PIV system. Therefore, this work aims to find an innovative way to predict the unavailable parameters mentioned earlier using existing flow fields.

Deep learning (DL) algorithms are rapidly developing and extensively used in various fields \cite{Pouyanfaretal2018, Morrisetal2023}. Recently, DL has been widely utilized in fluid dynamics, benefiting from its capability of highly nonlinear mapping \cite{Bruntonetal2020}. DL is of great interest in the following problems about turbulent flows: temporal flow data generation based on turbulent modeling \cite{Jiangetal2021, Duraisamyetal2019}; fluid flow simulation \cite{Vinuesa&Brunton2022}; reduced-order modeling \cite{Yousif&Lim2022, Eivazietal2022}; Prediction of turbulent flow based on the information from previous temporal data \cite{Yousifetal2022a, Yousifetal2022b, Fukamietal2019a}; super-resolution reconstruction of turbulent flow \cite{Fukamietal2019b, Yuetal2022, Yousifetal2022c, Yousifetal2021}. Another important application of DL in fluid dynamics is mapping parameters of the flow field, namely, mapping some parameters obtained from a fluid flow to the other different parameters, which can be expressed as

\begin{equation} \label{eqn:eq1}
f:P_a\rightarrow P_u,
\end{equation}

\noindent where $P_a$ is termed the set of available or already-existing parameters, $P_u$ is the set of unavailable parameters, and $f$ is the mapping rule taking the elements in $P_a$ to the one in $P_u$. The $f$ of note is the complex non-linear functions hidden in deep learning neural network, which are learned by the DL model through training. Guastoni {\it et al.} \cite{Guastonietal2021} applied convolutional neural network models to map the wall-shear-stress components and the wall pressure to the 2D instantaneous velocity-fluctuation fields at different wall-normal locations in a turbulent open-channel flow. Jouybari {\it et al.} \cite{Jouybarietal2021} designed a multilayer perceptron type neural network to get the mapping of various rough surface statistics to equivalent sand-grain height. Lee {\it et al.} \cite{Leeetal2022} proposed a transfer learning method based on empirical correlations to predict the drag force on the statistics of rough surfaces. Moreover, the mapping that takes local wall-shear stresses and wall pressure fluctuations to local heat flux was proved to be possible by Kim and Lee \cite{Kim&Lee2020}.

Furthermore, generative adversarial networks (GAN) proposed by Goodfellow {\it et al.} \cite{Goodfellowetal2020} have been used in various fields, e.g., the work of Shamsolmoali {\it et al.} \cite{Shamsolmoalietal2021} for synthetic image generation. Other powerful network based on GAN are super-resolution generative adversarial network (SRGAN) \cite{Ledigetal2017}, and the variant of SRGAN, i.e., enhanced SRGAN (ESRGAN) \cite{Wangetal2019}. Recently, GAN-based DL models have been rapidly implemented to solve turbulent flow problems, particularly resolution reconstruction problems. Benefiting from deeper layers and special loss functions (e.g., perceptual loss), SRGAN and ESRGAN show better performance on the resolution reconstruction of flow fields than basic other DL models \cite{Dengetal2019}. Moreover, Yousif {\it et al.} \cite{Yousifetal2022d} proposed a novel 2D3DGAN based on the mechanism of ESRGAN, which could reconstruct the 3D turbulent flow fields from 2D velocity fields. Regarding flow field parameters mapping, Guemes {\it et al.} \cite{Guemesetal2021} evaluated the performance of SRGAN for reconstructing turbulent-flow quantities from coarse wall measurements. The results showed that the SRGAN could capture and rebuild the large-scale structures of the flow even for the most complicated cases. Undoubtedly, the GAN-based DL models also have the potential to tackle more turbulence-related problems.

In this article, we attempt to predict unavailable parameters ($P_u$) from existing available parameters ($P_a$) using a deep learning model that can help find a mapping rule between $P_u$ and $P_a$. A novel GAN-based model is proposed, termed mapping-generative adversarial network (M-GAN), and is applied to map available parameters to unavailable parameters. Two cases of fluid flows are used to examine the performance of M-GAN. The 2D Rayleigh-B{\'e}nard (RB) flow at Rayleigh number ($Ra$) =${10}^8$ is used as a demonstration, where $u$ and $v$ are regarded as available parameters; meanwhile, $T$ is the unavailable parameter needing to be predicted. Furthermore, turbulent channel flow data at $Re_\tau$ = 180, 395, and 550, are applied to test the M-GAN model sufficiently. 2D flow fields at $Re_\tau$ =180 and 550 are the data for training and testing for M-GAN, i.e., $v$, and $w$ are input data, and $u$ and $p$ are output data. Then channel flow at $Re_\tau$ =395, which has never participated in the training process, is used as an additional testing case to estimate the interpolation ability of M-GAN. Note that a label information generator (LIG) is combined with M-GAN to train and test the turbulent channel flow cases. The role of LIG is to help M-GAN decide to output $u$ or $p$ when the DL model can only receive the same information of $v$ and $w$. All the mentioned flow data are generated using DNS.

The remainder of this paper is written as follows. Section II presents the generation of flow data using DNS. The design of M-GAN is introduced in the methodology part (section III), where the training and testing procedures are also included. The results, including instantaneous contours and turbulence statistics, are plotted and discussed in section IV. Finally, section V summarizes the article and proposes potential future research based on this study.

\section{Data generation}\label{sec:TBL-Generation}

\subsection{2D Rayleigh-B{\'e}nard flow}
In the Rayleigh-B{\'e}nard flow case, the governing equations with incompressibility condition can be expressed as 
\begin{equation} \label{eqn:eq2}
\frac{\partial {\bf u}}{\partial t} + {\bf u} \cdot \nabla {\bf u} = -\frac{1}{\rho}\nabla p + {\nu} {\nabla^2} {\bf u} + {\bf F_b},
\end{equation}

\begin{equation} \label{eqn:eq3}
\nabla \cdot {\bf u} = 0,
\end{equation}

As for the temperature field, an advection-diffusion equation is applied as
\begin{equation} \label{eqn:eq4}
{\frac{\partial {T}}{\partial t} + {\bf u} \cdot \nabla {T}} = {{\kappa}{\nabla^2}{T}},
\end{equation}

\noindent where ${\bf u}$, $p$, $\rho$, $T$, $\nu$, $\kappa$, and $t$ represent velocity vector, pressure, density, temperature, kinematic viscosity, thermal diffusivity, and time, respectively. The Boussinesq approximation is applied for this flow and the body force ${\bf F}_b$ is taken to only depend linearly on the temperature and to be in the direction of gravity. Besides, the possible dependencies of density, viscosity, and thermal diffusivity on temperature are ignored, so these parameters are considered constant.

The simulation is carried out by performing DNS using a well-validated second-order finite-difference code \cite{VerziccoandOrlandi1996, Zhuetal2018}. One important control parameter of Rayleigh-B{\'e}nard flow is the Rayleigh number, i.e., $Ra$=$\alpha g\Delta L^3/(\nu\kappa)$, where $\alpha$ is the thermal expansion coefficient, $g$ is the acceleration of gravity, and $\Delta$ is the temperature difference between the upper and bottom surface with a depth $L$. The no-slip and constant temperature boundary conditions are applied for the bottom and top plates, and periodic boundary condition is assigned to the horizontal direction. The details of this case are listed in Table~\ref{tab:Table1}. More information on the simulation of Rayleigh-B{\'e}nard flow can be found in these papers \cite{Zhuetal2018, Zhuetal2018a}.

\begin{table}
  \begin{center}
\scalebox{1.0}{
\begin{tabular}{ccccccccc} \hline\hline
&$Ra$~ ~&~~$N_x\times N_y$~ ~&~~$\mathrm{\Gamma}$~ ~&~~$Pr$~ ~&~~$Nu$~ ~&~~$\Delta t^+$&  \\ \hline
&$10^8$~ ~&~~$512 \times 256$~~&~~$2$~~&~~$1$~~&~~$26.1$~~&~~$0.1$&   \\
 \hline\hline
\end{tabular}}
  \caption{Simulation parameters of Rayleigh-B{\'e}nard flow. $N_x$ and $N_y$ are the grid resolution in the horizontal and vertical directions. $\mathrm{\Gamma}$ is the aspect ratio, $\mathrm{\Gamma}$=$W/L$, where $W$ and $L$ are the width and depth of the domain. $Pr$ and $Nu$ are the Prandtl number and Nusselt number. $\Delta t^+$ is the dimensionless time step of the simulation. The superscript $``+"$ represents that the quantity is nondimensionalized by $u_\tau$ and $\nu$.}
  \label{tab:Table1}
  \end{center}
\end{table}

In this case, 9000 snapshots are used as training data, while 3000 snapshots are used to test the trained model. To save the computational expense, all the data are interpolated to reduce the resolution from 512$\times$256 to 256$\times$128.

\subsection{Turbulent channel flow}
In the turbulent channel flow case, the momentum equation for an incompressible viscous fluid is expressed as
\begin{equation} \label{eqn:eq5}
\frac{\partial {\bf u}}{\partial t} + {\bf u} \cdot \nabla {\bf u} = -\frac{1}{\rho}\nabla p + {\nu} {\nabla^2} {\bf u} 
.
\end{equation}

The open-source CFD finite-volume code OpenFOAM-5.0x is applied to perform the DNS calculation. As mentioned earlier, the current study has used three different friction Reynolds numbers, i.e., $Re_\tau$ = $u_\tau \delta/\nu$ = 180, 395, and 550, where $u_\tau$ is the friction velocity, and $\delta$ is half of the channel height. The detailed simulation parameters of each $Re_\tau$ are listed in Table 2. The periodic boundary condition is used in the streamwise ($x$) and spanwise ($z$) directions. The no-slip boundary condition is applied to the upper and lower walls of the channel. The obtained turbulence statistics have been validated by comparing them with the results from Kim {\it et al.}\cite{Kimetal1987} and Moser {\it et al.}\cite{Moseretal1999}.

\begin{table}
  \begin{center}
\scalebox{1.0}{
\begin{tabular}{cccccccccc} \hline\hline
&$Re_\tau$~ ~&~~$L_x\times L_y \times L_z$~ ~&~~$N_x\times N_y \times N_z$~ ~&~~$\Delta x^+$~ ~&~~$\Delta z^+$~ ~&~~$\Delta y_w^+$~ ~&~~$\Delta y_c^+$~ ~&~~$\Delta t^+$&  \\ \hline
&$180$~ ~&~~$4\pi\delta\times2\delta\times2\pi\delta$~~&~~$256 \times 128 \times 256$~~&~~$8.831$~~&~~$4.415$~~&~~$0.63$~~&~~$4.68$~~&~~$0.113$&   \\
&$395$~ ~&~~$4\pi\delta\times2\delta\times2\pi\delta$~~&~~$385 \times 257 \times 385$~~&~~$12.553$~~&~~$6.277$~~&~~$0.541$~~&~~$5.115$~~&~~$0.023$& \\
&$550$~ ~&~~$4\pi\delta\times2\delta\times2\pi\delta$~~&~~$512 \times 336 \times 512$~~&~~$13.492$~~&~~$6.746$~~&~~$0.401$~~&~~$5.995$~~&~~$0.030$& \\ 
\hline\hline
\end{tabular}}
  \caption{Simulations parameters of two turbulent channel flows. $L$ and $N$ are the domain dimension and the number of grids. The superscript $``+"$ represents that the quantity is nondimensionalized by $u_\tau$ and $\nu$. $\Delta y_w^+$ and $\Delta y_c^+$ are the spacing near the wall and at the center of the channel.}
  \label{tab:Table2}
  \end{center}
\end{table}

We use the pressure implicit split operator algorithm to solve the coupled pressure momentum system. We discretize the convective fluxes using a second-order accurate linear upwind scheme. Additionally, all the discretization schemes employed in the simulations, including convective fluxes, have second-order accuracy. We maintain the maximum Courant–Friedrichs–Lewy (CFL) number below 1 during the simulations to ensure stability.

Training and testing data of a $y-z$ plane from the 3D channel flow domain are collected for both $Re_\tau$ = 180 and 550. 10,000 snapshots of flow velocity and pressure fields are obtained. The data of the flow at $Re_\tau$ =395 and 550 are interpolated to match the grid size of the DNS data of the flow at $Re_\tau$ = 180, which is 128 and 256 in $y$ and $z$ directions, respectively. By applying this data processing technique, we can ensure that the flow data at all three Reynolds numbers are suitable inputs for the model.

\begin{figure}
\centering 
\includegraphics[angle=0, trim=0 0 0 0, width=0.7\textwidth]{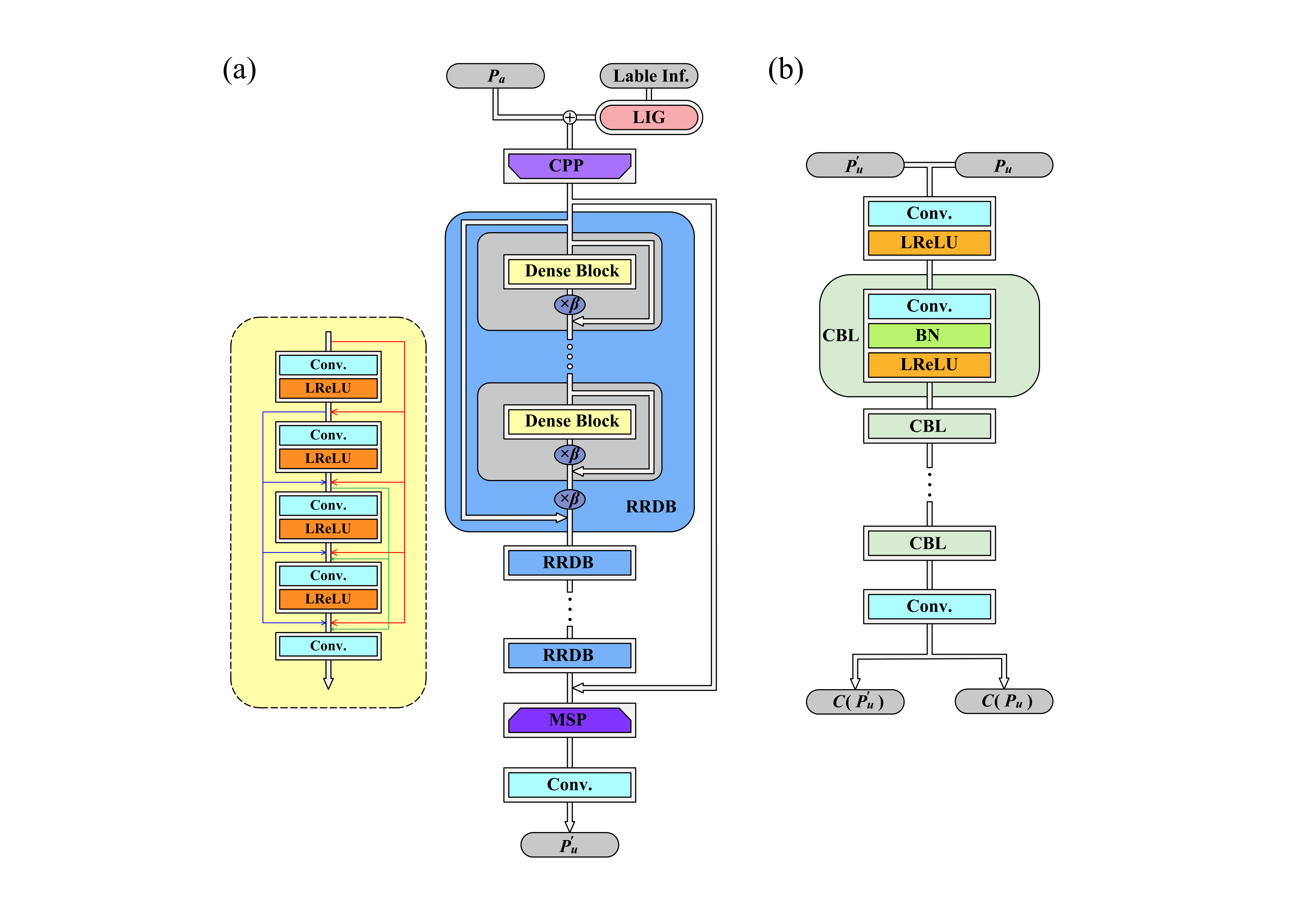}
\caption[]{Architecture of M-GAN: (a) generator and (b) discriminator.}
\label{fig:1-M-GAN}
\end{figure}

\section{Methodology}\label{sec:Method}

\subsection{M-GAN}

GAN, developed by Goodfellow {\it et al.} \cite{Goodfellowetal2020}, has been successfully applied to tackle the image generation problem. Subsequently, based on the traditional GAN, Ledig {\it et al.} \cite{Ledigetal2017} and Wang {\it et al.} \cite{Wangetal2019} proposed SRGAN and ESRGAN, respectively. SRGAN and ESRGAN have excellent performance in recovering high-resolution images from multiple images. In the turbulence field, SRGAN and ESRGAN have proven their capability to reconstruct the super-resolution data of turbulent flows \cite{Yuetal2022, Yousifetal2022c, Yousifetal2021}. Considering the previous great mapping ability of ESRGAN for resolution reconstruction, this study intends to develop the M-GAN to map some flow field parameters to other parameters.

Like the traditional GAN, M-GAN comprises two parts, i.e., generator ($G$) and discriminator ($D$). As shown in Fig.~\ref{fig:1-M-GAN} (a), $G$ mainly contains deep convolutional neural networks (CNNs)-based multi-layers termed residual in residual dense blocks (RRDBs) \cite{Zhangetal2021}. The sub-elements of RRDBs are dense blocks (DBs) consisting of convolutional and leaky ReLU activation function layers. Skip connection is applied in RRDBs and DBs to avoid vanishing gradient. In addition, $G$ comprises a convolution-pooling part (CPP) following the input and a multiscale part (MSP) preceding the output. The CPP module includes several convolutional and max pooling layers that extract input data information and compress the data size. Conversely, the MSP module uses upsampling layers and convolutional layers with various-sized filters to extract information from the output of RRDBs and construct the output data. More details of CPP and MSP are listed in Table~\ref{tab:CCP} and Table~\ref{tab:MSP}, respectively. Besides, the label information generator (LIG) is added to the input of $G$. LIG is used in specific situations, such as $v$ and $w$ are the input, and $u$ or $p$ are the output. Note that LIG can provide information to $G$ for determining which parameters to output. Table~\ref{tab:LIG} shows the detailed structure of LIG. The label is passed into a dense layer to generate label information. After reshaping, the label information is concatenated to $p_a$ and input into the next process. Notably, the use of LIG is optional. Sometimes LIG can be neglected when the outputting only includes one parameter in $P_u$. Fig.~\ref{fig:1-M-GAN} (b) shows the detailed structure of $D$. The Major component of $D$ is CBL, including three different layers, i.e., convolutional, batch normalization, and LReLU activation function layers.

\begin{table} 
\begin{center}
~\caption{CCP structure.}
\scalebox{1.0}{
\begin{tabular}{ c c }
\hline\hline
 Type of layers 			      & Shape 		           \\ \hline
 Input ($P_a$)		          &  (128, 256, 2) 			   \\ 
 Conv2D.(3, 3) 		          &  (128, 256, 56)		        \\  
 MaxPooling(2, 2) 		      &  (64, 128, 56)		        \\
 Conv2D.(3, 3) 		          &  (64, 128, 56)		        \\  
 MaxPooling(2, 2) 		      &  (32, 64, 56)		        \\
 concat(Output of last layer and LIG)       &  (32, 64, 64)       \\  \hline\hline
\end{tabular}} \label{tab:CCP}
\end{center}
\end{table}

\begin{table} 
\begin{center}
~\caption{MSP architecture.}
\scalebox{1.0}{
\begin{tabular}{ c c c }
\hline\hline
 First branch 			& Second branch 			&  Third branch 	\\ \hline
 Conv2D.(3, 3) 		&  Conv2D.(5, 5) 			&  Conv2D.(7, 7) 		\\ 
 UpSampling(2, 2)		&  UpSampling(2, 2)		&  UpSampling(2, 2)	 \\  
 Conv2D.(3, 3) 		&  Conv2D.(5, 5) 			&  Conv2D.(7, 7) 		\\ 
 LeakyReLU			&  LeakyReLU				&  LeakyReLU		\\  
 UpSampling(2, 2)		&  UpSampling(2, 2)		&  UpSampling(2, 2)	 \\ 
 Conv2D.(3, 3) 		&  Conv2D.(5, 5) 			&  Conv2D.(7, 7) 		\\ 
 LeakyReLU			&  LeakyReLU				&  LeakyReLU		\\  \hline
 \multicolumn{3}{c} {Add (first branch,  second branch, third branch) } \\ \hline \hline
\end{tabular}} \label{tab:MSP}
\end{center}
\end{table}

\begin{table} 
\begin{center}
~\caption{LIG structure.}
\scalebox{1.0}{
\begin{tabular}{ c c }
\hline\hline
 Type of layers 			& Shape 		  \\ \hline
 Input (label inf.) 		&  (1) 			   \\ 
 Dense		            &  (32×64×8)		\\  
 Reshape                 &  (32, 64, 8)    \\  \hline\hline
\end{tabular}} \label{tab:LIG}
\end{center}
\end{table}

The overall algorithm is described as follows. First, $P_a$ and label information (here, we assume to have the label information) are input to $G$. The generated unavailable parameters ($p_u^{'}$) are obtained from the output of $G$. Then, both $P_u$ (real one) and $P_u^{'}$ (generated one) are fed into $D$. The non-transformed discriminator value $D_{Ra}$ is calculated from the output of $D$, which is formulated as

\begin{equation} \label{eqn:eq6}
D_{Ra} (P_u , P_u^{'} ) = \sigma \left(C \left(P_u \right)  - \mathbb{E}_{P_u^{'}} \left[C ( P_u^{'}) \right]\right),
\end{equation}

\begin{equation} \label{eqn:eq7}
D_{Ra} (P_u^{'} , P_u ) = \sigma \left(C \left(P_u^{'}\right)  - \mathbb{E}_{P_u} \left[C ( P_u) \right]\right),
\end{equation}

\noindent where $\sigma$ is the sigmoid function, and $\mathbb{E}$ is the average calculating operator. Similar to ESRGAN, $D$ is designed to predict the probability that $P_u $ is relatively more realistic than $P_u^{'}$. Using equations~\ref{eqn:eq6} and~\ref{eqn:eq7}, the discriminator loss $L_D^{Ra}$ and adversarial loss $L_G^{Ra}$ are calculated. The formulation of the discriminator loss ($\mathcal{L}_D$) is expressed as

\begin{equation} \label{eqn:eq8}
\mathcal{L}_D = L_D^{Ra} = -\mathbb{E}_{P_u} \left[ {\rm log} (D_{Ra} (P_u , P_u^{'} )) \right] - \mathbb{E}_{P_u^{'}} \left[ {\rm log} (1 - D_{Ra} (P_u^{'} , P_u )) \right].
\end{equation}
Besides, the adversarial loss is expressed as

\begin{equation} \label{eqn:eq9}
L_G^{Ra} = -\mathbb{E}_{P_u} \left[ {\rm log} (1 - D_{Ra} (P_u , P_u^{'} )) \right] - \mathbb{E}_{P_u^{'}} \left[ {\rm log} (D_{Ra} (P_u^{'} , P_u )) \right].
\end{equation}

As mentioned before, $D$ predicts the probability that the $P_u$ is relatively more realistic than the $P_u^{'}$ in the training process. When $P_u$ is more realistic than $P_u^{'}$, $D_{Ra}(P_u,P_u^{'} )$ tends to be 1 and $D_{Ra}(P_u^{'},P_u )$ tends to be 0. Thus, $L_D^{Ra}$ will decrease and tend to be 0. Inversely, $G$ plays a role in generating more realistic $P_u^{'}$ tending to be similar to $P_u$. In this situation, $D_{Ra}(P_u,P_u^{'} )$ will decrease from 1 to 0, and $D_{Ra}(P_u^{'},P_u )$ will increase from 0 to 1, which makes $L_G^{Ra}$ tend to be 0. The above description is the adversarial process between $G$ and $D$. In this process, $G$ and $D$ compete with each other. Meanwhile, they are also promoted mutually.

\begin{figure}
\centering 
\includegraphics[angle=0, trim=0 0 0 0, width=0.8\textwidth]{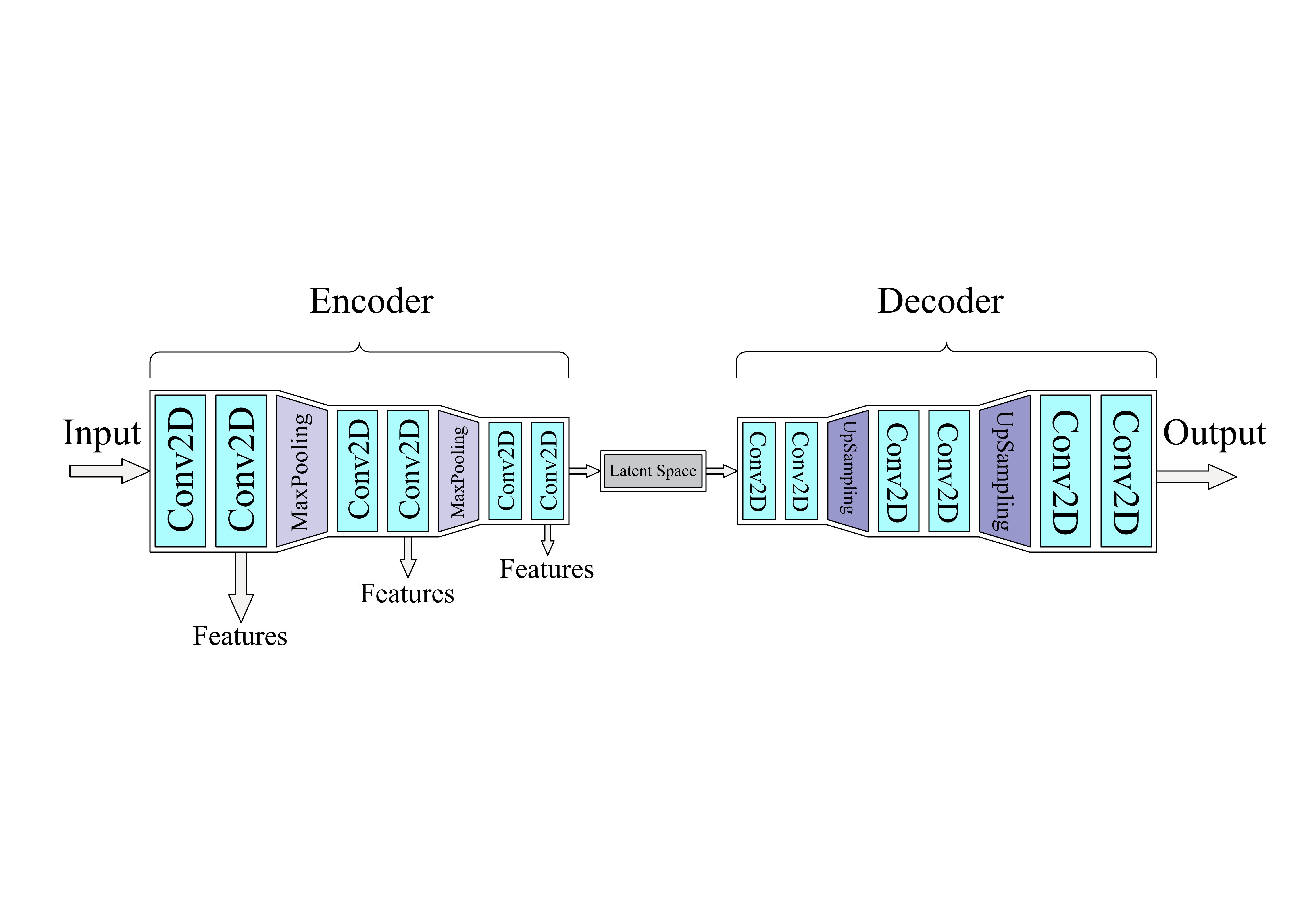}
\caption[]{Architecture of the AE-FE.}
\label{fig:2-AE-FE}
\end{figure}

In addition to adversarial loss, the loss function of G ($\mathcal{L}_G$) contains some extra parts, i.e., pixel loss ($L_{pixel}$), perceptual loss ($L_{perceptual}$). $L_{pixel}$ is the pixel-based error, which is calculated by comparing the value difference between $P_u $ and $P_u^{'}$. Different from $L_{pixel}$, $L_{perceptual}$  is the error between the features of $P_u$ and $P_u^{'}$, which are extracted by the feature extractor (FE). Regarding the original ESRGAN, VGG19 \cite{Simonyan&Zisserman2015}, a very deep convolutional network, is used as FE. However, VGG19 is designed to process image data with three channels, i.e., RGB. In other words, VGG19 can only receive the flow data with three components like $u$, $v$, and $w$. In this study, $P_u$ and $P_u^{'}$ only contains one component, so applying VGG19 to extract features is unsuitable. This study develops an autoencoder-based DL algorithm as a FE (AE-FE) to replace VGG19. FE is a pre-trained model using $P_u$ data. As shown in Fig.~\ref{fig:2-AE-FE}, after the $P_u$ are input the AE-FE, various features data with different sizes are output from several convolutional layers. The above extra loss terms are computed using the Mean-squared error (MSE). The total loss function of the $G$ is expressed as

\begin{equation} \label{eqn:eq10}
\mathcal{L}_G = L_G^{Ra} +\lambda L_{pixel} + L_{perceptual},
\end{equation}

\noindent where, $\lambda$ is the coefficient used to balance the magnitude of various loss terms, which is set to be 1000.

This study uses the open-source library TensorFlow 2.3.0 to implement the DL model. The customized sample Python code for the proposed M-GAN is available on the web page (\url{https://fluids.pusan.ac.kr/fluids/65416/subview.do}).

\subsection{Training and testing of M-GAN}

\begin{figure}
\centering 
\includegraphics[angle=0, trim=0 0 0 0, width=0.8\textwidth]{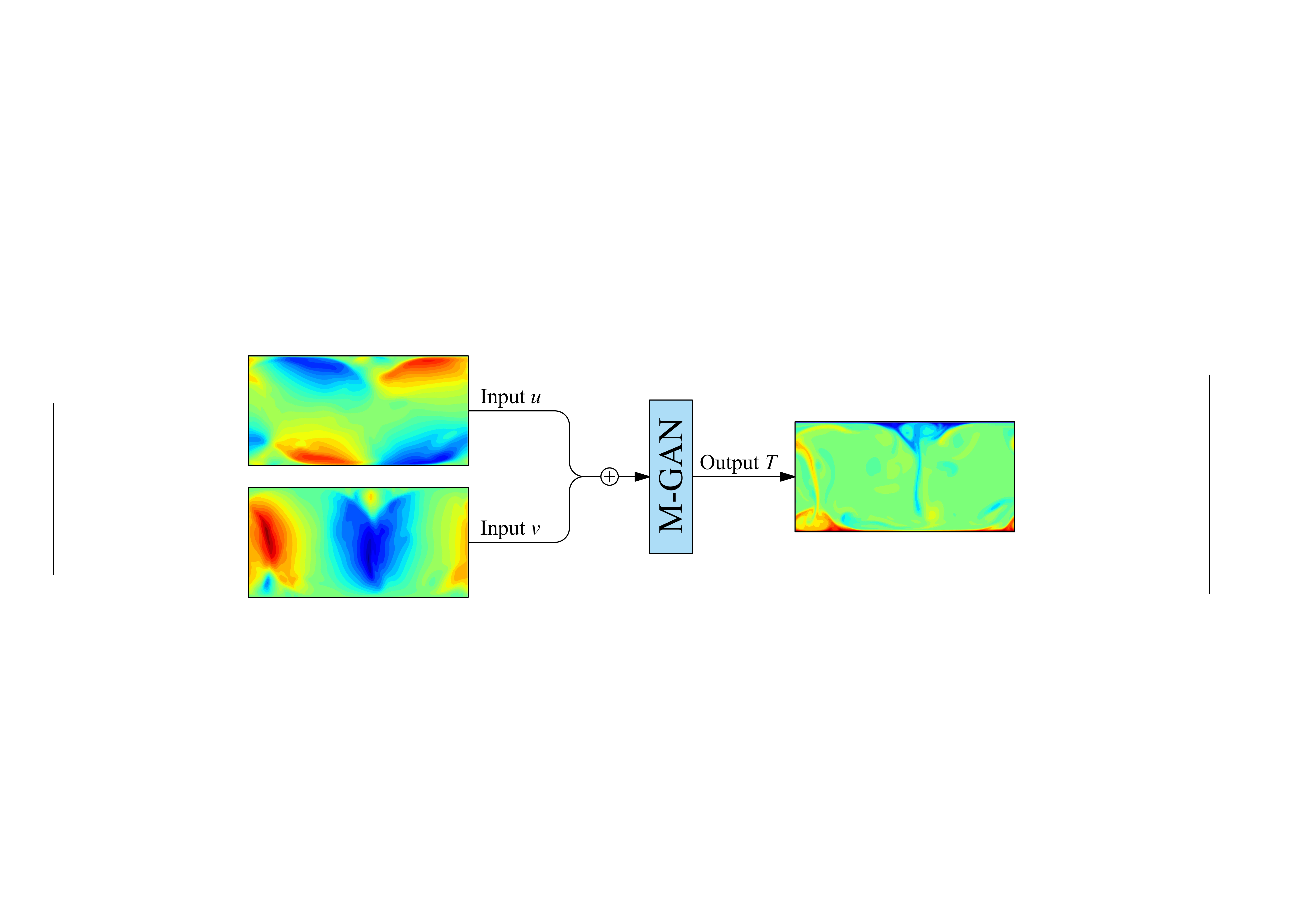}
\caption[]{Schematic of the 2D Rayleigh-B{\'e}nard flow case.}
\label{fig:RBcase}
\end{figure}

For the training process of the 2D Rayleigh-B{\'e}nard flow case, the input data, i.e., $P_a$ are velocity components $u$ and $v$. Meanwhile, the target data, i.e., $P_u$ is temperature $T$. Firstly, $u$ and $v$ are input into $G$ to output $P_u^{'}$, with a batch size of 16. Then, generator loss is calculated using $P_u$ and $P_u^{'}$ based on equation~\ref{eqn:eq10}. The discriminator loss is calculated by equation~\ref{eqn:eq8} through $P_u$ and $P_u^{'}$. After computing the losses, the optimization algorithm updates the weights of both $G$ and $D$ to minimize losses during the training period. The adaptive moment estimation (Adam) algorithm is used in this study for optimization \cite{Kingma&Ba2017}. In addition, the target data, $P_u$, is normalized using the min-max normalization function with a range of 0 to 1 to enhance the training performance. After training, the $u$ and $v$ values in the test dataset are fed into trained M-GAN to generate predicted $T$ ($P_u^{'}$), which is depicted in Fig.~\ref{fig:RBcase}.

\begin{figure}
\centering 
\includegraphics[angle=0, trim=0 0 0 0, width=0.8\textwidth]{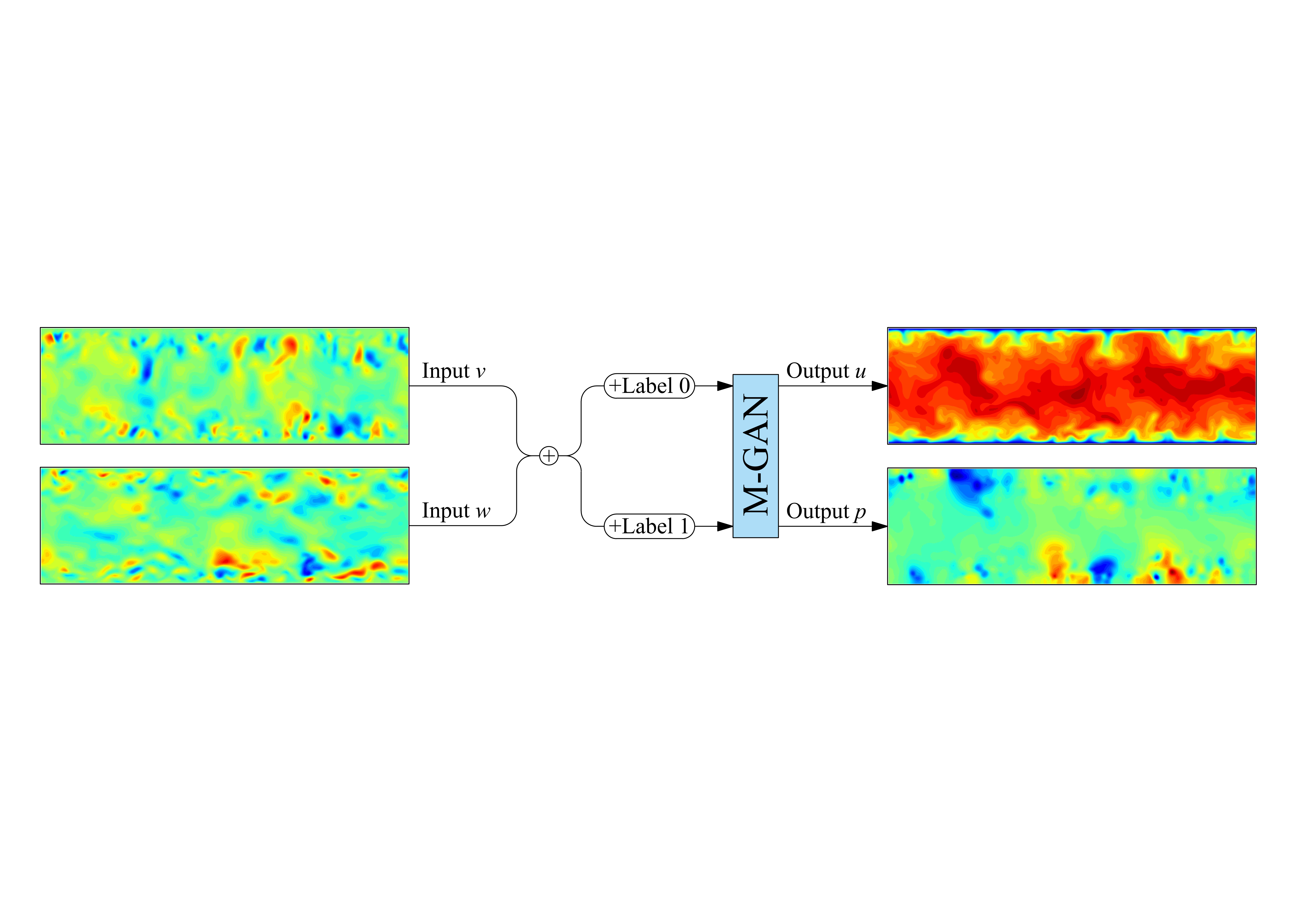}
\caption[]{Schematic of the Channel flow case.}
\label{fig:CFcase180550}
\end{figure}

In the training process of the turbulent channel flow case, the input data comprise wall-normal velocity $v$ and spanwise velocity $w$, while the target data consist of the streamwise velocity $u$ and pressure $p$. For predicting $u$, a label of 0 is inputted into LIG, whereas a label of 1 is used for predicting $p$. In contrast to the 2D Rayleigh-B{\'e}nard flow case, we train the M-GAN using both channel flow datasets at $Re_\tau$ = 180 and 550 together. The training details and methods are similar to the 2D Rayleigh-B{\'e}nard flow case mentioned above, so the detailed explanation is neglected here. As shown in Fig.~\ref{fig:CFcase180550}, we input $v$ and $w$ from channel flow and the corresponding label into the trained M-GAN to predict the corresponding $u$ or $p$.

Here, the channel flow data at $Re_\tau$ = 395 mentioned above never attend the training process of M-GAN, and they are used to evaluate the interpolation ability of M-GAN. The $v$ and $w$ from channel flow at $Re_\tau$ = 395 and label are inputted into the M-GAN trained by channel flows at $Re_\tau$ = 180 and 550 to predict the $u$ or $p$. Here, if the M-GAN trained by channel flows at $Re_\tau$ = 180 and 550 could also predict the unavailable parameters of channel flow at $Re_\tau$ = 395 from its $v$ and $w$, it could be proved that the proposed M-GAN has good interpolation ability. In other words, the M-GAN might be appropriate for all the channel flows at a range of Reynolds numbers ($Re_\tau$ = 180 to 550) to predict their unavailable parameters. The verification of the instantaneous and statistical results from all the above cases will be presented in due course.

\section{Results and discussion}
\subsection{2D Rayleigh-B{\'e}nard flow}

\begin{figure}
\centering 
\includegraphics[angle=0, trim=0 0 0 0, width=0.8\textwidth]{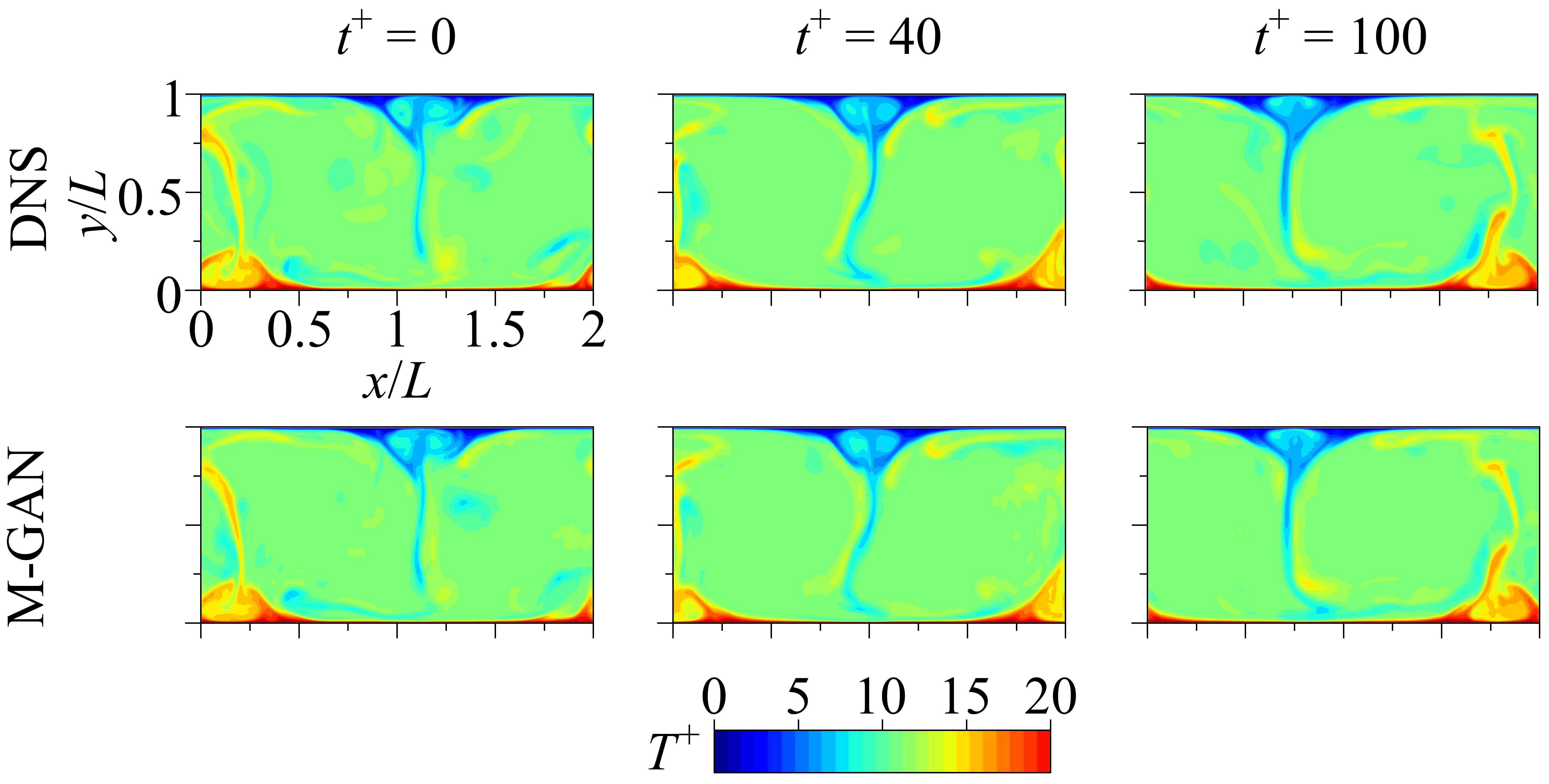}
\caption[]{Predicted instantaneous temperature field of the 2D Rayleigh-B{\'e}nard flow case.}
\label{fig:RBC}
\end{figure}

The capability of M-GAN is examined in this sub-section for the case of 2D Rayleigh-B{\'e}nard flow. Fig.~\ref{fig:RBC} shows the predicted instantaneous temperature field from the mapping of velocity components using M-GAN. The defined parameters: $t^+$ is the dimensionless time, and $T^+$is the dimensionless temperature calculated from $T^+$ = $T/T_\tau$, where $T_\tau$ = $-\kappa \partial_y\langle T\rangle_{x,t}|_{y=0}/u_\tau$ is a characteristic temperature scale like the $u_\tau$ for the velocity \cite{Yaglom1979}. From these instantaneous temperature contours, we can find that all the detailed features of predicted results using M-GAN are consistent with the corresponding DNS data, even if the intense large-scale rolls (LSRs) keep moving along x direction with time going.

\begin{figure}
\centering 
\includegraphics[angle=0, trim=0 0 0 0, width=0.8\textwidth]{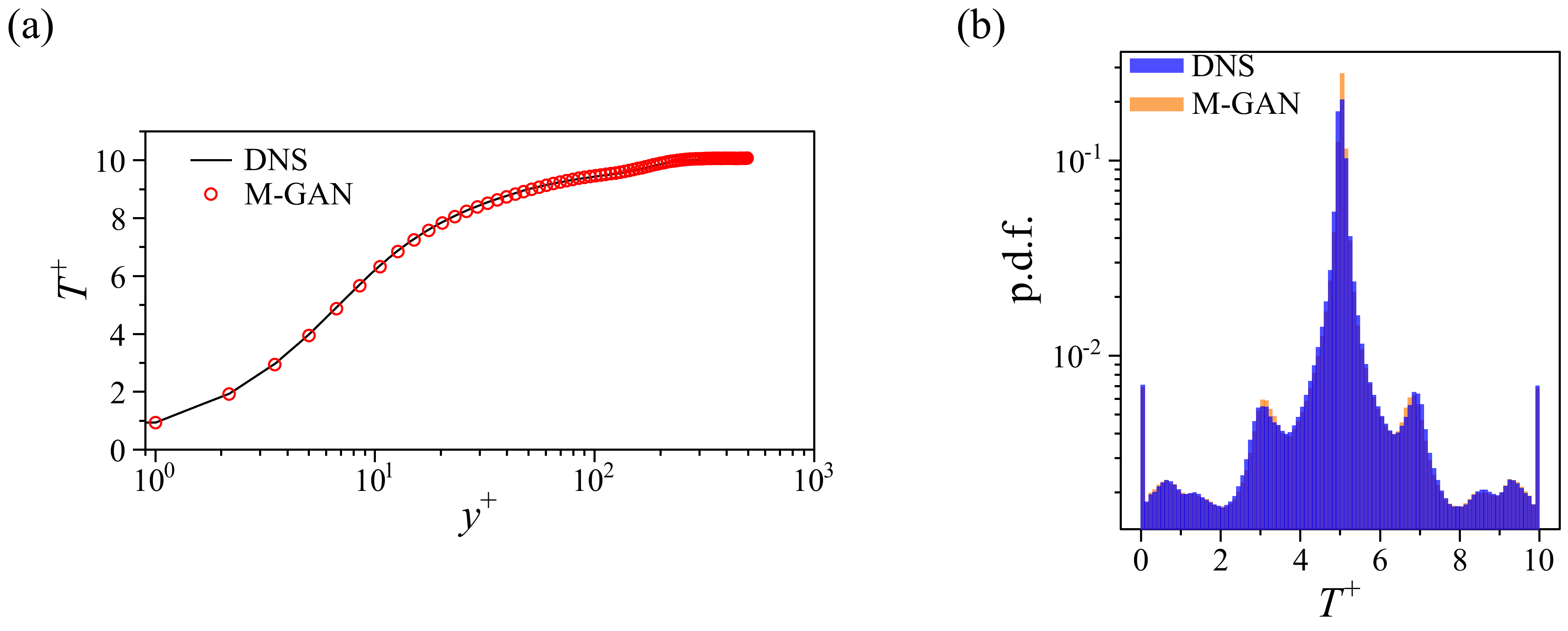}
\caption[]{Statistical results of the 2D Rayleigh-B{\'e}nard flow case: (a) average temperature profile; (b) probability density function of temperature field.}
\label{fig:RBST}
\end{figure}

Moreover, the mapping ability of M-GAN is further validated by plotting and comparing the statistical results from the predicted and DNS data. The average temperature profile along the wall unit $y^+$ and the probability density function (PDF)  are plotted in Fig.~\ref{fig:RBST}. All the predicted results show commendable agreement with the results obtained from DNS data indicating that M-GAN can map the available parameters $u$ and $v$ to unavailable parameter $T$ of the 2D Rayleigh-B{\'e}nard flow. 

Although the results show that M-GAN can work well for the RB flow case, the RB flow case still cannot sufficiently verify the mapping ability of M-GAs, because Rayleigh-B{\'e}nard flow does not have very complex and chaotic behavior at a low Rayleigh number ($Ra$=$10^8$). Thus, the next sub-section will illustrate a more comprehensive verification using turbulent channel flow data.
\subsection{Turbulent channel flow}
\begin{figure}
\centering 
\includegraphics[angle=0, trim=0 0 0 0, width=0.8\textwidth]{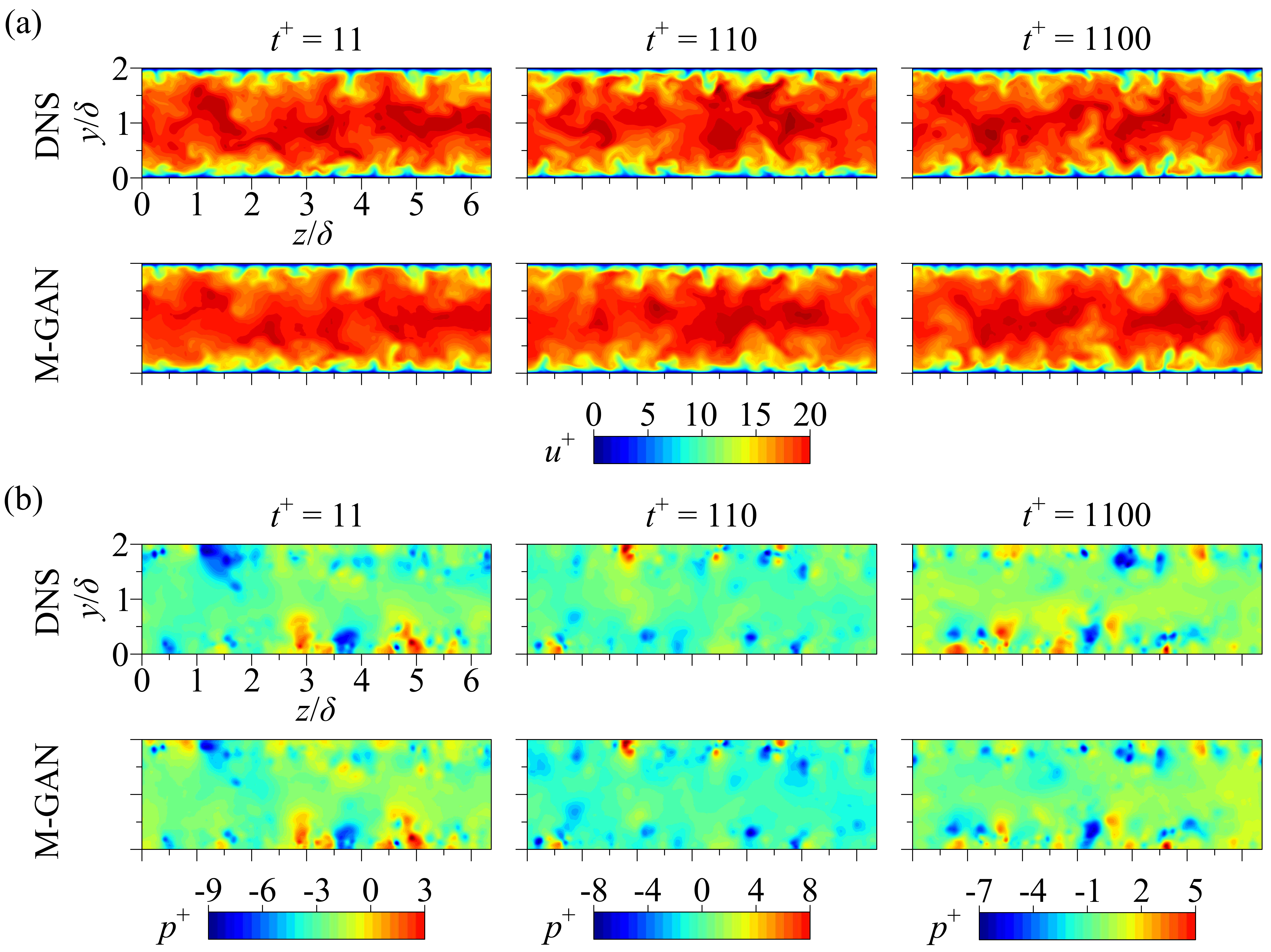}
\caption[]{Predicted instantaneous streamwise velocity and pressure fields of the turbulent channel flow at $Re_\tau$ = 180.}
\label{fig:CFC180}
\end{figure}

\begin{figure}
\centering 
\includegraphics[angle=0, trim=0 0 0 0, width=0.8\textwidth]{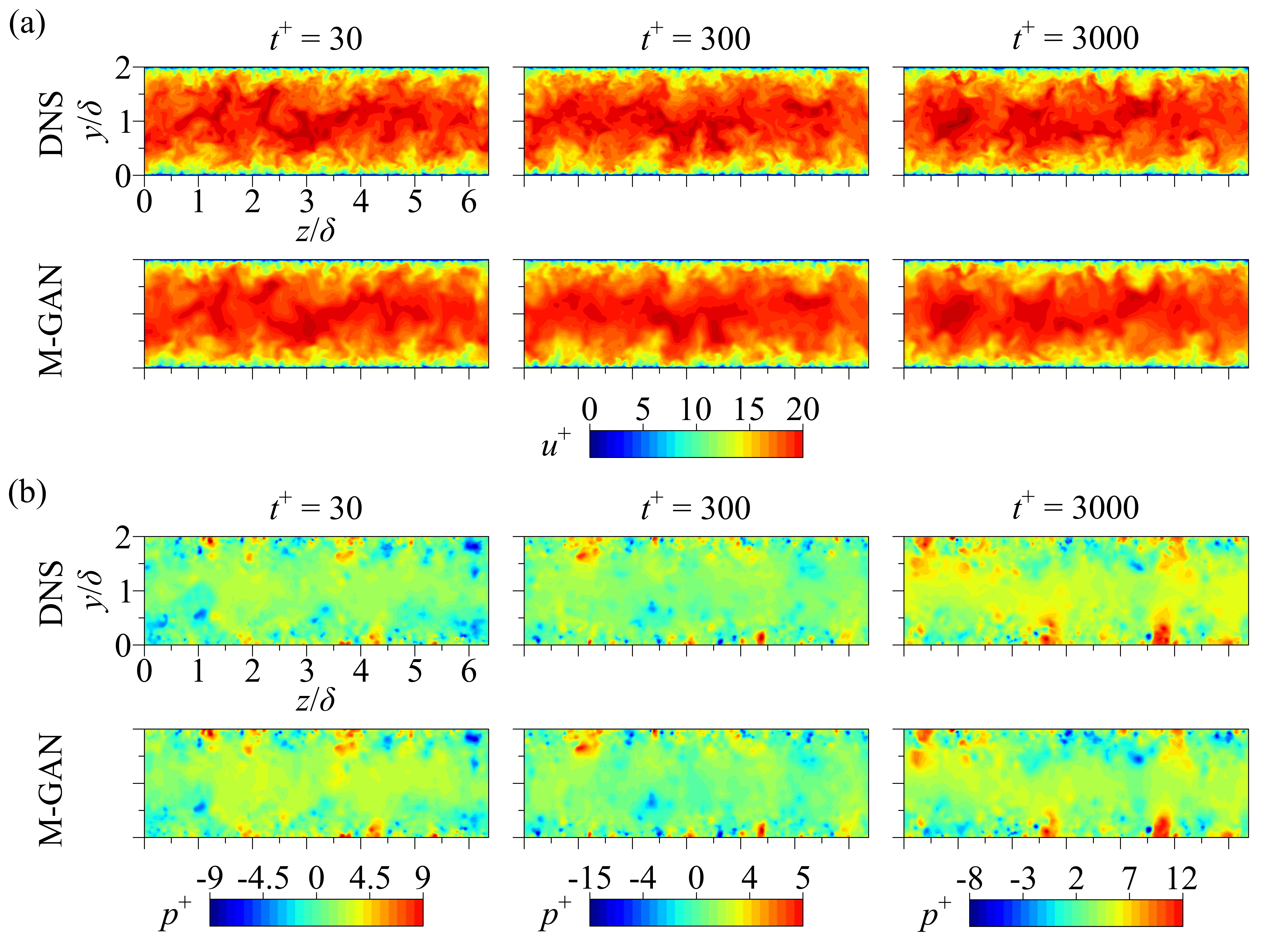}
\caption[]{Predicted instantaneous streamwise velocity and pressure fields of the turbulent channel flow at $Re_\tau$ = 550.}
\label{fig:CFC550}
\end{figure}

\begin{figure}
\centering 
\includegraphics[angle=0, trim=0 0 0 0, width=0.8\textwidth]{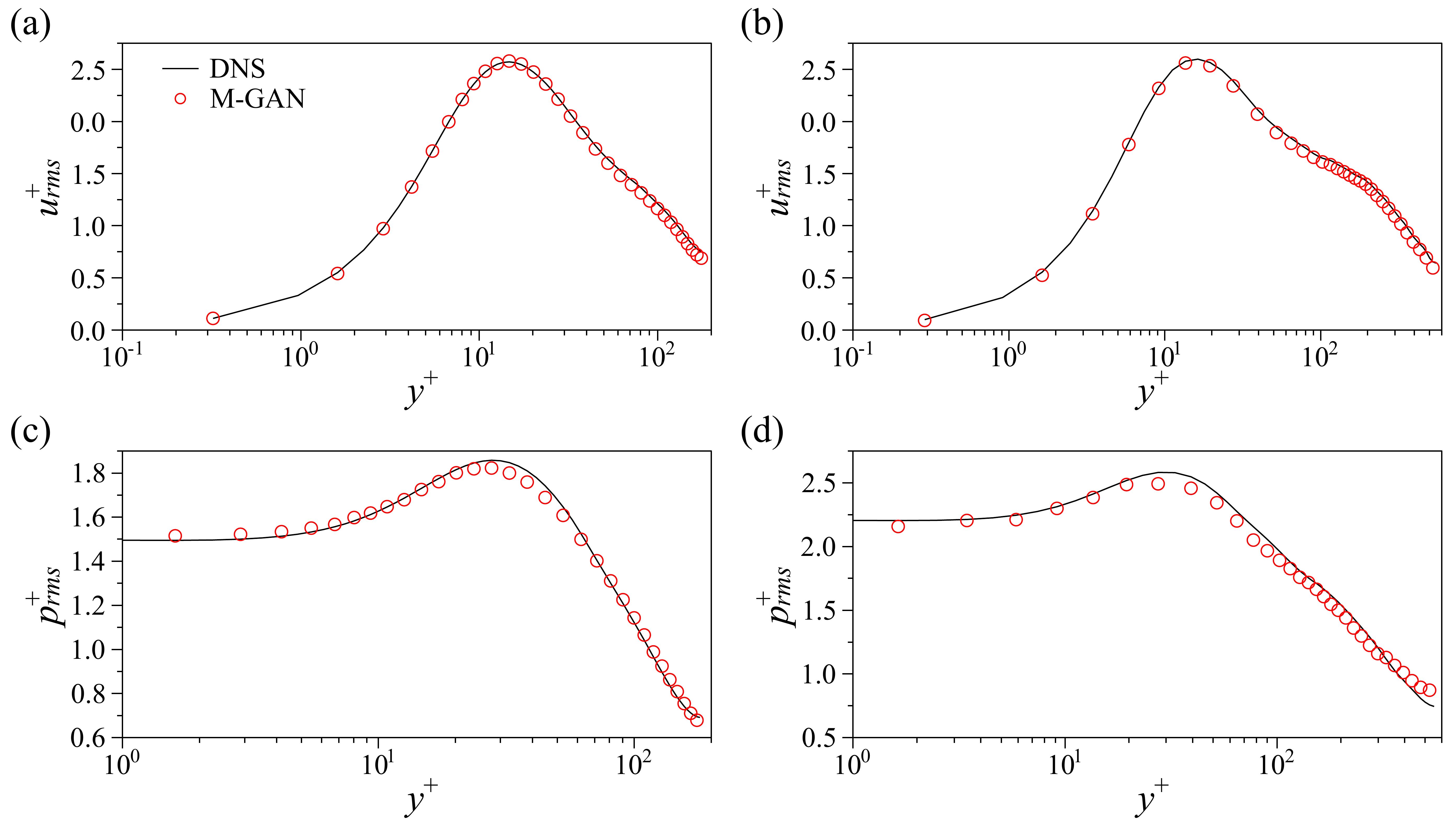}
\caption[]{Root-mean-square profiles of streamwise velocity  and pressure fluctuations: (a) and (c) are $u_{rms}^+$ and $p_{rms}^+$ of the channel flow at $Re_\tau$ = 180; (b) and (d) are $u_{rms}^+$ and $p_{rms}^+$ of the channel flow at $Re_\tau$ = 550.}
\label{fig:CFRMS180550}
\end{figure}

In this sub-section, the mapping capability of M-GAN for taking the available parameters $v$ and $w$ to unavailable parameters $u$ and $p$ is examined. Fig.~\ref{fig:CFC180} and Fig.~\ref{fig:CFC550} show the predicted instantaneous streamwise velocity and pressure fields of the turbulent channel flow at $Re_\tau$ = 180 and 550. The predicted results of both cases agree well with related the DNS results, where most fluctuating features of predicted results are consistent with the DNS data. Through observation by comparing the results based on different parameters and $Re_\tau$, we can find that the results of channel flow at $Re_\tau$ = 180 have more accuracy as compared to the results from the results of the flow at $Re_\tau$ = 550. Besides, when comparing the results from the same channel flow case, the predicted $u$ field is better than the results from $p$. Fig.~\ref{fig:CFRMS180550} also indicates this observation, where root-mean-square (RMS) velocity and pressure fluctuations ($u_{rms}^+$ and $p_{rms}^+$) are plotted. The reason for that could be attributed to the fact that pressure fields are more chaotic than streamwise velocity fields, and channel flow at $Re_\tau$ = 550 is more complicated than the one at $Re_\tau$ = 180.

\begin{figure}
\centering 
\includegraphics[angle=0, trim=0 0 0 0, width=0.8\textwidth]{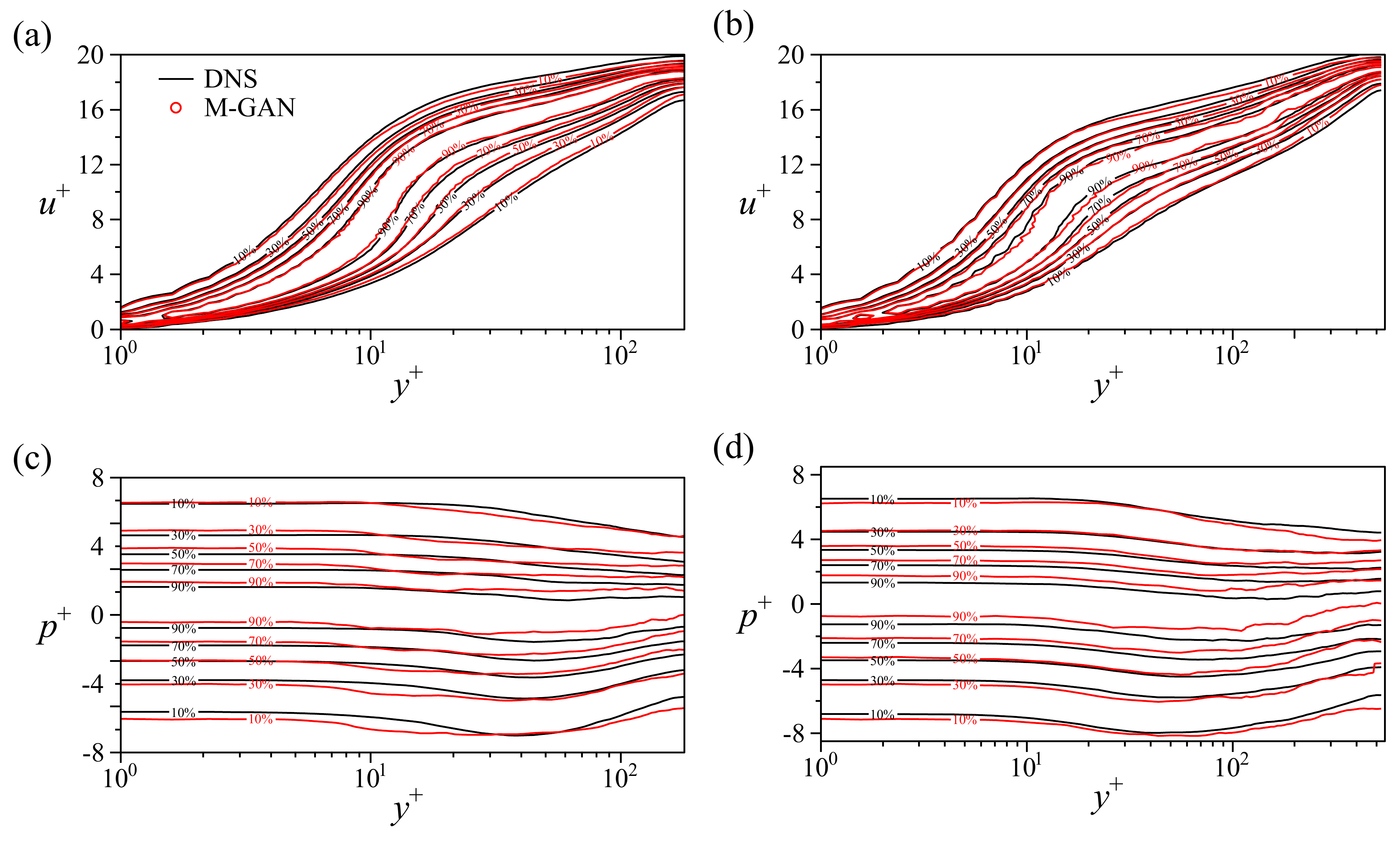}
\caption[]{Probability density functions of the streamwise velocity and pressure fields of channel flows: (a) and (c) are PDF($u$) and PDF($p$) of the channel flow at $Re_\tau$ = 180; (b) and (d) are PDF($u$) and PDF($p$) of the channel flow at $Re_\tau$ = 550. The isoline levels are in the range of 10\% – 90\% of the maximum PDF with an increment of 20\%.}
\label{fig:CFPDF180550}
\end{figure}

Figure~\ref{fig:CFPDF180550} shows the probability density function (PDF) plots of the $u$ and $p$ fields. Note that a larger probability means that the parameter magnitude appears more frequently and dominates the flow field to a greater extent. The PDF of predicted velocity and pressure fields can basically match the reference results. However, the results obtained from channel flow at $Re_\tau$ = 550 show less accuracy than the results from channel flow at $Re_\tau$ = 180. At the same time, the deviation shown in PDF($p$) is much larger than the one in PDF($u$), which is consistent with the previous discussion in the first paragraph of this subsection.

\begin{figure}
\centering 
\includegraphics[angle=0, trim=0 0 0 0, width=0.7\textwidth]{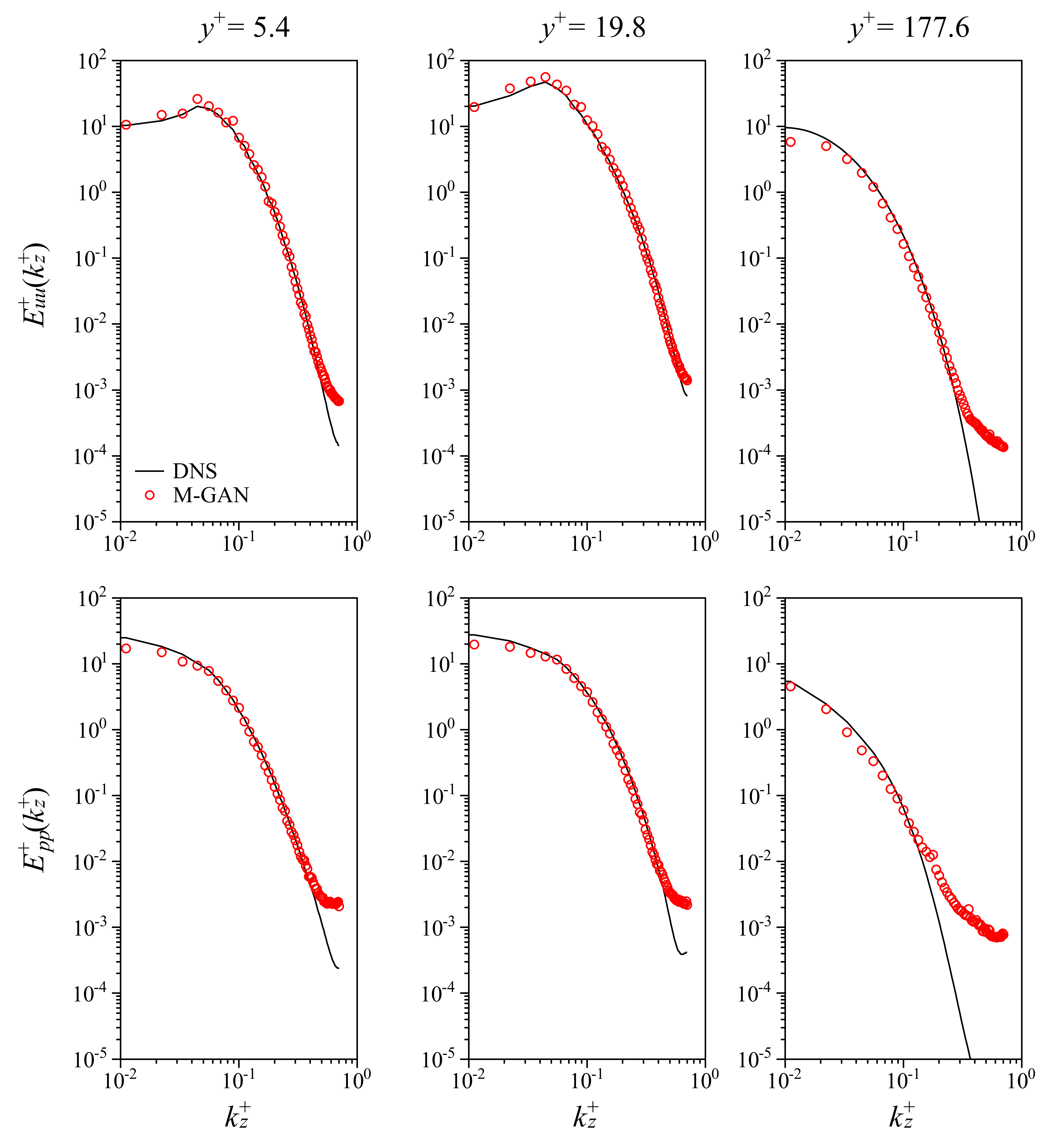}
\caption[]{Spanwise energy spectra of the predicted streamwise velocity and pressure from channel flow at $Re_\tau$ = 180.}
\label{fig:CFSPC180}
\end{figure}

\begin{figure}
\centering 
\includegraphics[angle=0, trim=0 0 0 0, width= 0.7\textwidth]{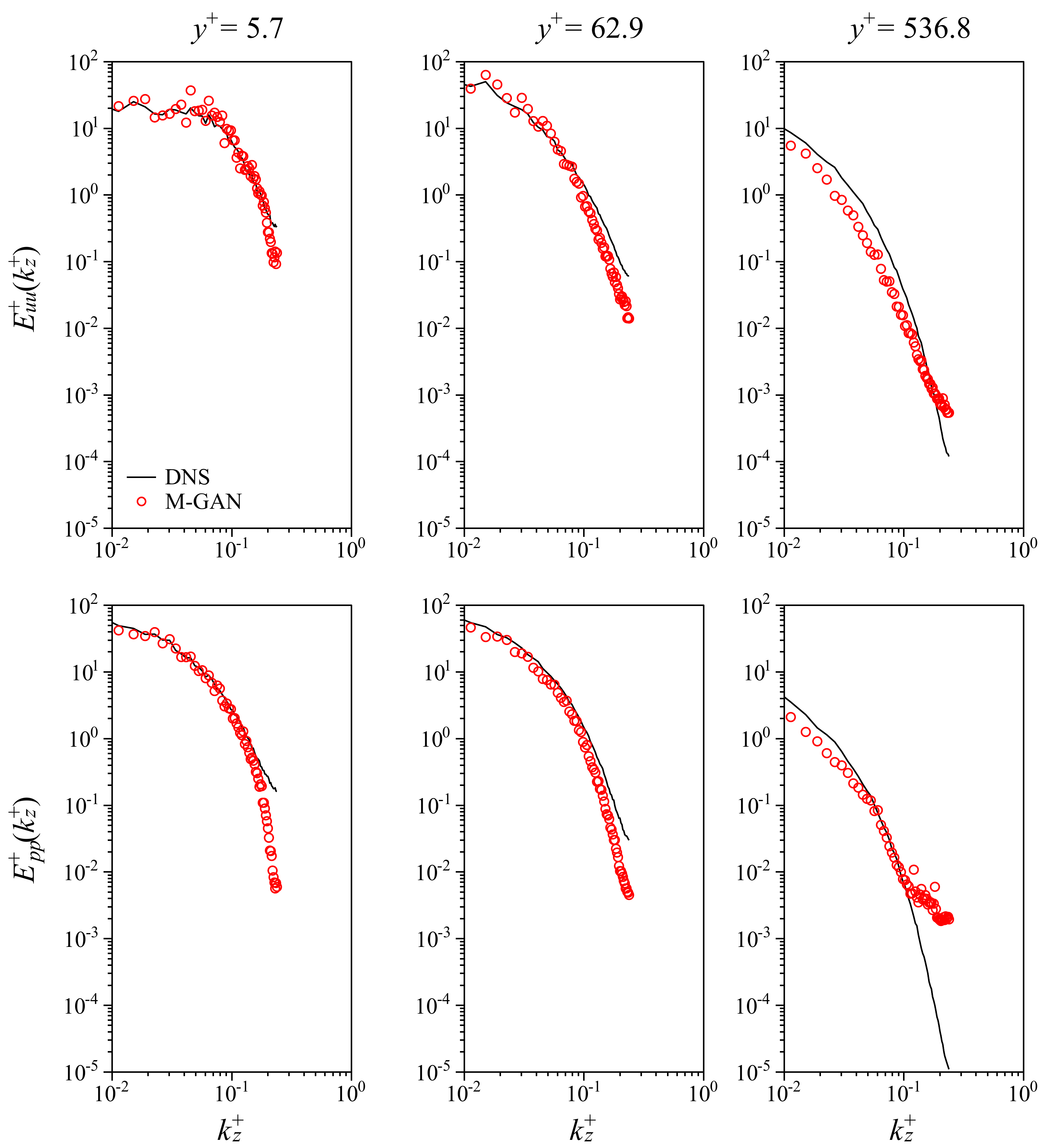}
\caption[]{Spanwise energy spectra of the predicted streamwise velocity and pressure from channel flow at $Re_\tau$ = 550}
\label{fig:CFSPC550}
\end{figure}

In Fig.~\ref{fig:CFSPC180} and Fig.~\ref{fig:CFSPC550}, the spanwise energy spectrum of $u$ and $p$ ($E_{uu}^+ (k_z^+)$ and $E_{pp}^+ (k_z^+)$) at various wall distances are plotted to evaluate the ability of M-GAN to predict the $u$ and $p$ fields with realistic behavior. In these two figures, the results of $E_{uu}^+ (k_z^+)$ have more accuracy than the results of $E_{pp}^+ (k_z^+)$). Besides, with the $y^+$ increasing, the deviation of the energy spectrum also increases. This consistent with previously discussed instantaneous contours and RMS results. Another interesting thing observed from the energy spectrum results is that in the flow at $Re_\tau$ = 180, the predicted energy in high wavenumbers is higher than the DNS results. On the contrary, the predicted energy in high wavenumbers is mostly lower than the DNS results for the channel flow case at $Re_\tau$ = 550.

With the above discussion, we can observe that the M-GAN can predict the unavailable parameters from the corresponding available data based on flow data used for training. However, the interpolation ability of the model still needs to be further investigated, In other words, whether the model can work in the channel flow case at a Reynolds number between $Re_\tau$ = 180 and 550 or not. Regarding this issue, a channel flow at $Re_\tau$ = 395, which never joins the training, is used for the interpolation ability test. Same as the previous operation, $v$ and $w$ are regarded as available data to predict $u$ and $p$ using pre-trained M-GAN trained by the data of channel flows at $Re_\tau$ = 180 and 550. 

\begin{figure}
\centering 
\includegraphics[angle=0, trim=0 0 0 0, width=0.8\textwidth]{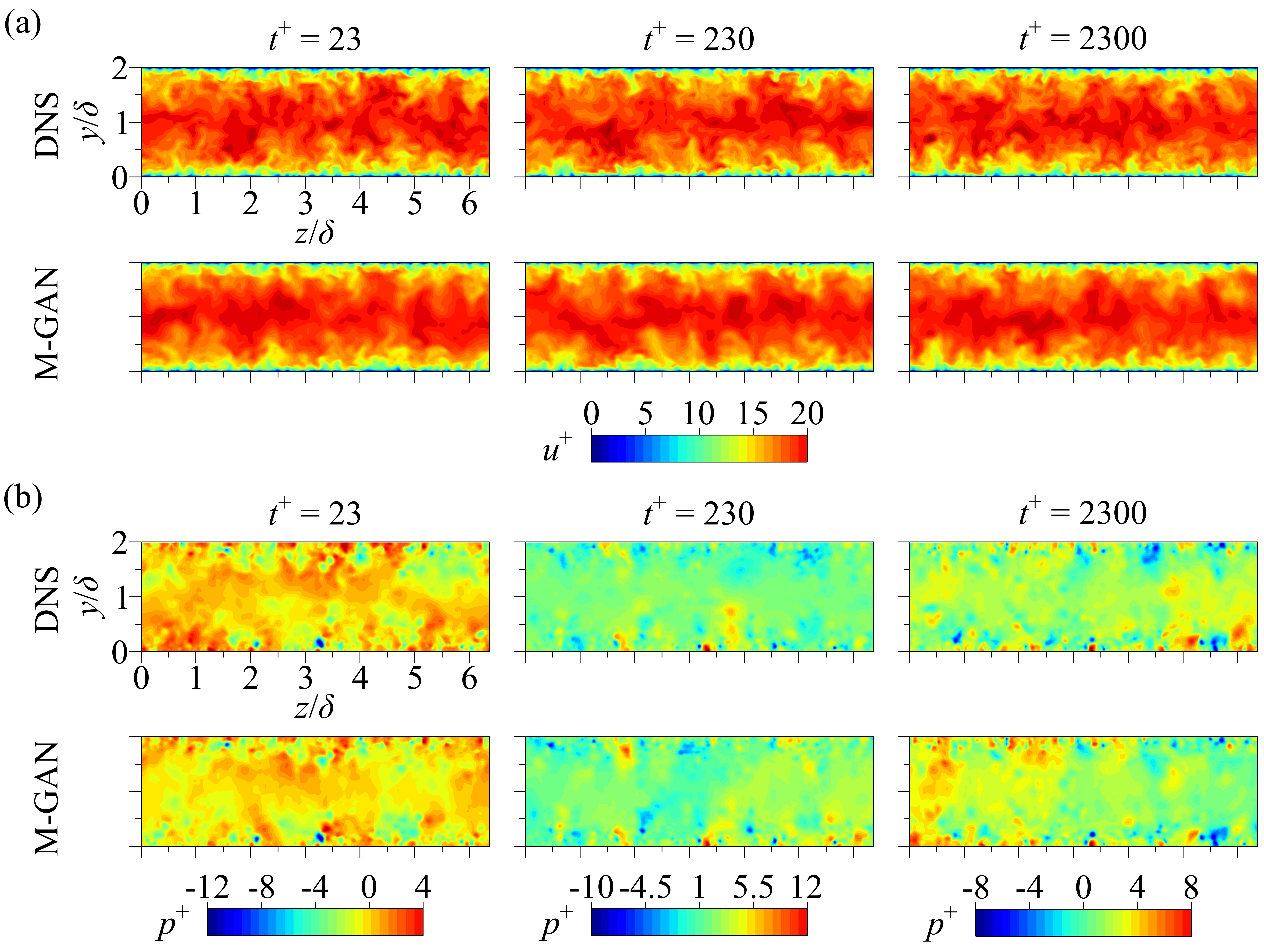}
\caption[]{Predicted instantaneous streamwise velocity (a) and pressure (b) fields of the turbulent channel flow at $Re_\tau$ = 395.}
\label{fig:CFC395}
\end{figure}

\begin{figure}
\centering 
\includegraphics[angle=0, trim=0 0 0 0, width=0.8\textwidth]{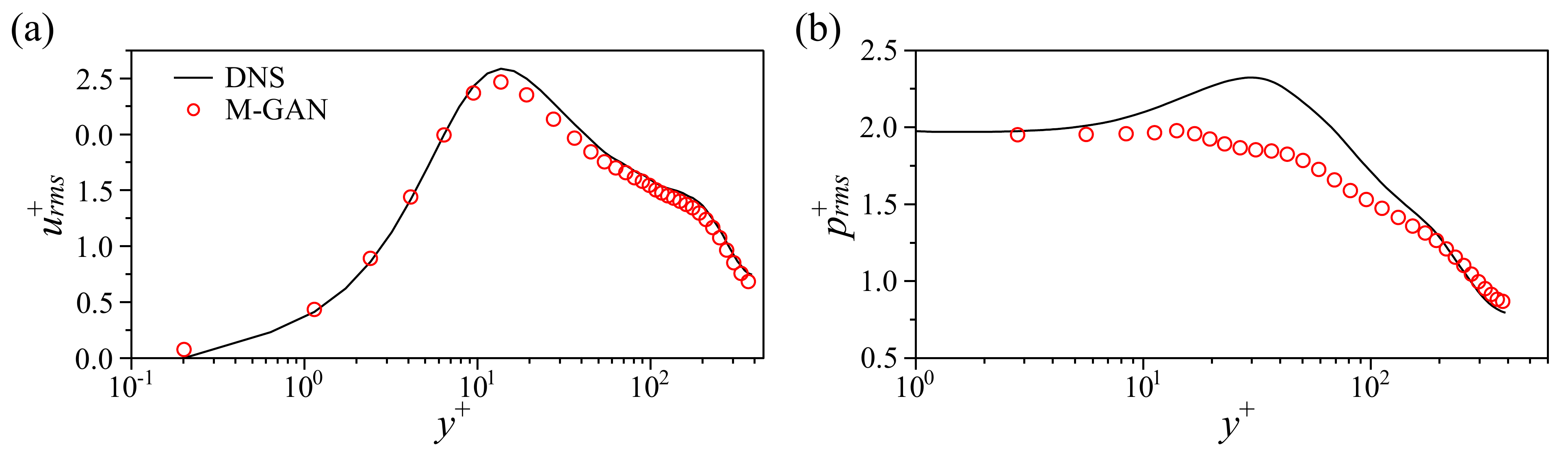}
\caption[]{Root-mean-square profile of the streamwise velocity (a) and pressure (b) fluctuations of the  turbulent channel flow at $Re_\tau$ = 395.}
\label{fig:CFRMS395}
\end{figure}

Fig.~\ref{fig:CFC395} shows the predicted instantaneous streamwise and pressure fields for the flow at $Re_\tau$ = 395 . Moreover, Fig.~\ref{fig:CFRMS395} shows the statistical results of RMS velocity and pressure fluctuations. Although there are larger errors than in previous results, especially the $p$ field cannot be predicted well, the results still are acceptable. Generally, most of the features of flow fields can be constructed and $u_{rms}^+$ has a relatively good agreement with the DNS result. This indicates that the proposed M-GAN has a good interpolation ability to work successfully in a case different from the training data set. In other words, M-GAN can learn the mapping function between $P_a$ and $P_u$.

\section{Conclusions}\label{sec:Conclusions}

We proposed an efficient deep learning-based model, namely, M-GAN to predict unavailable flow parameters from existing ones. Two cases of classical fluid flows were used to validate the performance and interpolation ability of M-GAN.

First, the 2D Rayleigh-B{\'e}nard flow was used as a demonstrating case to show the capability of M-GAN for mapping available parameters $u$ and $v$ to unavailable parameters $T$. In this case, the results indicated that the M-GAN could indeed predict the temperature field of the 2D Rayleigh-B{\'e}nard flow from its velocity field.

Then, two turbulent channel flows at $Re_\tau$ = 180 and 550 were used together to train M-GAN and validate the predicting performance of M-GAN. In this case, wall-normal velocity $v$ and spanwise velocity $w$ were regarded as available data and passed into M-GAN to predict unavailable parameters that are streamwise velocity $u$ and pressure $p$. Here, LIG was applied to help M-GAN decide which parameter ($u$ or $p$) should be output when the input layer received the same $v$ and $w$. The prediction's instant contours and turbulence statistics were exhibited and compared to DNS data. The results illustrated that M-GAN successfully predicted $u$ and $p$ from the corresponding $v$ and $w$ in both turbulent channel flow cases. However, the pressure field and the flow fields of the channel flow at $Re_\tau$ = 550 were relatively more difficult to be predicted because of their more chaotic turbulence characteristics.
Finally, turbulent channel flow at $Re_\tau$ = 395, which was never used to train the M-GAN, was employed to test the interpolation ability of M-GAN. As the results showed, the pre-trained M-GAN, which was based on the training data set of channel flows at $Re_\tau$ = 180 and 550, also could predict the unavailable parameters $u$ and $p$ from $v$ and $w$ from channel flow data at $Re_\tau$ = 395. Thus, it was proved that M-GAN learned the mapping law from $v$ and $w$ to $u$ and $p$ of the channel flows at a range of Reynolds numbers ($Re_\tau$ = 180 to 550). In other words, M-GAN showed a good interpolation ability for a specific range of Reynolds numbers.
In conclusion, this article has shown that M-GAN can predict unavailable flow parameters from existing ones, i.e., velocity fields. Besides, with good interpolation ability, M-GAN can work, even for turbulent flows with a range of Reynolds number.

\section*{Acknowledgments}
This work was supported by 'Human Resources Program in Energy Technology' of the Korea Institute of Energy Technology Evaluation and Planning (KETEP), granted financial resource from the Ministry of Trade, Industry \& Energy, Republic of Korea (no. 20214000000140). In addition, this work was supported by the National Research Foundation of Korea (NRF) grant funded by the Korea government (MSIP) (no. 2019R1I1A3A01058576). This work was also supported by the National Supercomputing Center with supercomputing resources including technical support (KSC-2022-CRE-0282).

\section*{Data Availability}
The data that supports the findings of this study are available within this article.
% The \nocite command causes all entries in a bibliography to be printed out
% whether or not they are actually referenced in the text. This is appropriate
% for the sample file to show the different styles of references, but authors
% most likely will not want to use it.
\nocite{1}

%Bibliography
\bibliographystyle{unsrt}  
\bibliography{my-bib}% Produces the bibliography via BibTeX.

\end{document}